\documentclass{article}

\usepackage{graphicx}
\usepackage{bm}

\newcommand{\be}{\begin{equation}}
\newcommand{\ee}{\end{equation}}

\begin{document}

\begin{flushright}
Liverpool Preprint: LTH 612\\
Helsinki preprint HIP-2003-62/TH \\
 \today
 \end{flushright}
  
\vspace{5mm}
\begin{center}
{\LARGE \bf Excited B mesons from the lattice}\\[10mm] 
{\large\it UKQCD Collaboration}\\[3mm]

{\bf A.M. Green, J. Koponen} \\
{Department of Physical Sciences and Helsinki Institute of
Physics\\
P.O. Box 64, FIN--00014 University of Helsinki, Finland}\\[3mm]
{\bf C. McNeile, C. Michael, G. Thompson}\\
{Department of Mathematical Sciences, University of Liverpool,
 L69 3BX, UK}\\[2mm]
 \end{center}
 
\date{\today}

\begin{abstract} We determine the energies of the excited states of a 
heavy-light meson $Q\bar{q}$, with a static heavy quark  and light quark
with mass approximately that of the strange quark  from both quenched
lattices and  with dynamical fermions. We are able to explore the 
energies of orbital excitations up to L=3,  the spin-orbit splitting up
to L=2 and the first radial excitation.   These $b \bar{s}$  mesons  
will be  very narrow if their  mass is less than 5775 MeV  --- the $BK$
threshold.  We investigate this in detail and present evidence that the
scalar  meson (L=1) will be very narrow and that as many as 6 $b \bar{s}$
excited  states will have energies close to the $BK$ threshold and
should also be relatively narrow.

\end{abstract}

\section{Introduction}

The spectroscopy of excited $B$ and $D$ mesons is important for our
understanding of QCD. Moreover, as stressed recently by
Rosner~\cite{Rosner:2003cz,Gronau:1992ke}, $B^*$ states also have
important applications to CP studies of neutral $B$ mesons  by the
identification of their flavour ($b$ versus $\bar{b}$) 
through the decay chain $B^* \to B^0 \pi^{\pm}$. Hence narrow 
$B^*$ resonances will be valuable for this.

  In the heavy quark limit, the $\bar{Q} q$ meson, which we refer to as a
`$B$' meson, will be the `hydrogen  atom' of QCD. Since the meson is
made from non-identical quarks, charge conjugation  is not a good
quantum number.  States can be  labelled by  L$_{\pm}$, where the
coupling of the light quark spin to the orbital angular momentum  gives 
$j_q=L\pm {1 \over 2}$. In the heavy quark  limit these states will be
doubly degenerate since the heavy quark spin interaction can be 
neglected, so  the  P$_{-}$ state will have  $J^{P}=0^+,\ 1^+$ while
P$_{+}$ has $J^{P}=1^+,\ 2^+$, \textit{etc.}

 This spectrum has been studied comprehensively by lattice methods in
the quenched  approximation~\cite{M+P} with a rather coarse lattice
spacing. 
 With $N_f=2$ flavours of  dynamical quark, SESAM~\cite{Bali}  have
explored the P$_-$ excited state.  
 Lattice studies using NRQCD have also explored the heavy-light spectrum 
for $b$ quarks, mainly using quenched lattices~\cite{collins,Hein,lewis}.

 In the heavy quark effective theory, the leading order is just the 
static limit. The next correction will be of order $1/m_Q$ and will  be
relatively small for $b$ quarks, but larger for $c$ quarks. One way to 
predict the spectrum for $b$ quarks is to interpolate between charmed
states, where the experimental spectrum is  known, and the static  limit
obtained by lattice QCD assuming a dependence as $1/m_Q$.  Thus the
splittings among $B$ mesons should be approximately 0.33 of those  among
the corresponding $D$ mesons.

 The striking discovery that the $c\bar{s}$ states with $J^P=0^+$ and
$1^+$  have very narrow widths~\cite{expt} raises the question of whether
the  corresponding   $b\bar{s}$ states will also be narrow. The main 
reason for the narrow width of $D_s$ mesons is that the 
transition to $DK$  is not energetically allowed (for the 2317 MeV state)
or the state is  close to threshold (for the 2457 MeV state). Thus the only 
allowed hadronic decay proceeds via isospin-violation (since $m_d \not =
m_u$)  to $D_s$$\pi$ and will have a very small width. 
 Likewise, if the equivalent $b \bar{s}$ states are close to or  below
the $BK$ threshold, then they will be very  narrow. One of our main tasks
will be to determine the energy of these  excited states as accurately
as possible to check this.

 As well as exploring this issue of great interest to experiment, we
determine the  excited state spectrum  of the heavy-light system as
fully as possible. This will help  the construction of  phenomenological
models and  will shed light on questions such as  whether there is an
inversion of the level ordering (with L$_+$ lighter than L$_-$)  at
larger L or for radial excitations as   has been
predicted~\cite{HJ1,HJ2}.
We also compare with chiral models~\cite{chiral}.

 \section{Lattice evaluation}

 We investigate the heavy-light meson spectrum from lattice QCD using
static heavy quarks. Previous lattice studies have explored~\cite{M+P}
the full spectrum (\textit{i.e.}  S, P$_-$, P$_+$, D$_-$, D$_+$, F) in
quenched QCD. 
 There has also been a recent determination of the P$_-$  excitation energy
in full QCD~\cite{Bali}.  

 Here we present a range of different lattice studies: with different 
spatial volumes, lattice spacings and light quark masses --- see
Tables~\ref{tab_df},\ \ref{tab_qu}.
 We follow the all-to-all methods used in the static-light  lattice
study of Michael and  Peisa~\cite{M+P}.   Keeping their parameters, we
first use a larger  spatial size of lattice to check for finite size
effects --- Q1 vs. Q3. 
We are also able to correct the assignments of D$_+$ and D$_-$
states in their  work, see Q1 and Q2 in Table~\ref{tab_qu}.
 Our major study involves using lattice configurations~\cite{Allton176,
Allton202} which include $N_f=2$ flavours of  sea-quark, with two
different lattice spacings. We only use the unitary points, namely 
those with valence light quarks of the same mass as the sea quarks.  The
details are collected in Table~\ref{tab_df}.

\begin{table}
 \begin{tabular}{|l|ll|ll|}
 \hline                                                            
                & DF1           & DF2           & DF3           & DF4           \\   
 \hline
 $\beta$        & 5.2           & 5.2           & 5.2           & 5.2           \\
 $C_{SW}$       & 1.76          & 1.76          & 2.0171        & 2.0171        \\
 no.            & 20            & 78            & 20            & 40            \\
 Volume         &$12^3\times 24$&$16^3\times 24$&$16^3\times 32$&$16^3\times 32$\\
 $\kappa$       & 0.1395        & 0.1395        & 0.1350        & 0.1355        \\
 $r_0/a$        & 3.435         & 3.444         & 4.754         & 5.041         \\
 $r_0 m(0^{-+})$& 1.92(4)       & 1.94(3)       & 1.93(3)       & 1.48(3)       \\
 \hline                                                                         
 1S             &  3.00(5)      &  2.90(2)      &  3.68(7)      & 3.73(8)       \\
 2S             &  4.24(11)     &  4.10(5)      &  5.61(8)      & 5.60(14)      \\
 1P$_{-}$       &  4.01(6)      &  4.02(3)      &  4.71(8)      & 4.75(6)       \\
 2P$_{-}$       &  5.52(7)      &  5.57(5)      &  7.1(2)       & 7.38(9)       \\
 1P$_{+}$       &  4.18(11)     &  4.19(14)     &  5.4(3)       & 5.5(2)        \\
 2P$_{+}$       &  5.9(2)       &  5.57(5)      &  8.0(2)       & 8.35(14)      \\
 1D$_{-}$       &  5.32(12)     &  5.13(10)     &  6.6(2)       & 6.85(10)      \\
 2D$_{-}$       &  6.5(2)       &  6.35(14)     &  8.4(2)       & 8.9(2)        \\
 1D$_{+}$       &  5.73(8)      &  5.2(2)       &  7.05(14)     & 7.39(8)       \\
 2D$_{+}$       &  6.61(8)      &  6.7(3)       &  8.84(12)     & 8.99(7)       \\
 1D$_{+-}$      &  5.22(5)      &  5.17(4)      &  6.69(11)     & 7.22(6)       \\
 2D$_{+-}$      &  5.99(8)      &  6.06(10)     &  8.0(2)       & 8.47(10)      \\
 1F$_{+-}$      &  6.60(4)      &  6.25(4)      &  8.08(9)      & 7.94(12)      \\
 2F$_{+-}$      &  7.03(4)      &  6.97(3)      &  9.17(5)      & 9.53(8)       \\
 \hline   
\end{tabular}	    

 \caption{ Lattice results for the energies of $Q\bar{q}$ states in
units of $r_0$ for dynamical fermions with $N_f=2$. Here $r_0$ is taken to be
0.525(25) fm.}
 \label{tab_df}
\end{table}

\begin{table}
 \begin{tabular}{|l|l|l|l|}
 \hline   
                & Q1            & Q2             & Q3            \\   
 \hline
 $\beta$        & 5.7           & 5.7            & 5.7           \\
 $C_{SW}$       & 1.57          & 1.57           & 1.57          \\
 no.            & 20            & 20             & 20            \\
 Volume         &$12^3\times 24$&$12^3\times 24$ &$16^3\times 24$\\
 $\kappa$       & 0.14077       & 0.13843        & 0.14077       \\
 $r_0/a$        & 2.94          & 2.94           & 2.94          \\
 $r_0 m(0^{-+})$& 1.555(6)      & 2.164(6)       & 1.555(6)      \\
 \hline                                                   
 1S             & 2.57(2)       & 2.68(2)        & 2.555(12)     \\
 2S             & 3.74(3)       & 3.78(3)        & 3.70(2)       \\
 1P$_{-}$       & 3.57(13)      & 3.86(5)        & 3.62(10)      \\
 2P$_{-}$       & 5.1(2)        & 5.28(9)        & 5.0(2)        \\
 1P$_{+}$       & 3.7(2)        & 4.08(8)        & 3.82(6)       \\
 2P$_{+}$       & 5.0(2)        & 5.36(7)        & 5.0(2)        \\
 1D$_{-}$       & 4.80(10)      & 4.89(4)        & 4.67(7)       \\
 2D$_{-}$       & 5.7(2)        & 5.67(4)        & 5.60(11)      \\
 1D$_{+}$       & 4.8(2)        & 4.91(4)        & 4.98(5)       \\
 2D$_{+}$       & 5.8(3)        & 5.78(6)        & 5.69(5)       \\
 1D$_{+-}$      & 4.57(4)       & 4.64(3)        & 4.54(3)       \\
 2D$_{+-}$      & 5.37(10)      & 5.37(9)        & 5.29(6)       \\
 1F$_{+-}$      & 5.44(11)      & 5.60(7)        & 5.45(9)       \\
 2F$_{+-}$      & 6.04(13)      & 6.2(2)         & 6.04(7)       \\
 \hline                                                    
\end{tabular}	    

 \caption{ Lattice results for the energies of $Q\bar{q}$ states in
units of $r_0$ in the quenched case. Here results Q1, Q2 are from
Ref.~\cite{M+P}  with their D$_+$ and D$_-$ corrected. }
  \label{tab_qu}
\end{table}

 To extract mass values, we use operators with the appropriate 
representations of the cubic group (as described by \cite{M+P}) with
different degrees of non-locality.   We find that the D$_{+-}$ operator 
approximately gives the spin-average of the D$_-$ and D$_+$ levels, so
we can  interpret the F$_{+-}$ operator as representing an average of
the two F levels.  Our choice of operators  enables us to determine $N
\times N$ matrices of correlations for each  case, where $N$ can vary from
2 to 5. We then perform a fit to  these correlations over a suitable
$t$-range with a number of states  allowed. The requirement is then that
the  $\chi^2$ per degree of freedom is reasonable (not much  greater
than 1).
 We always use at least 2 states so that we have a reliable estimate of
the  ground state mass. To extract the first excited state as well, it
is preferable to use  at least a 3 state fit. We check that using  a
subset of our largest matrix of correlators, using different $t$-ranges,
 using one more or less state, \textit{etc.} gives stable results.

 To compare different lattice simulations, we form the dimensionless
combination  of $r_0 m$, where $m$ is a mass or energy, and where $r_0/a$
is determined relatively accurately  from the static quark potential.
 Our results are shown in Tables~\ref{tab_df},\ \ref{tab_qu}  and some
comparisons for  the P- and D-wave states are shown  in Figs.~1-3 versus
lattice spacing and versus quark mass.

 In order to relate our lattice results to experiment we have to 
discuss three different extrapolations.

 (i) {\em Finite Size Effects}. The lattice spatial volume should be
large enough.   There are several related criteria: the wavefunctions of
the heavy light  mesons should be small compared to the spatial size
$L_S$, the exchange of  the lightest particle (the pseudoscalar meson)
around the periodic boundary should be small  and mixing of the
heavy-light mesonic states with two body states (\textit{e.g.} $B\pi$
where the $\pi$ has  a low  momentum) should be small.

 We can estimate the size of the heavy-light mesons  from the
Bethe-Salpeter wavefunctions measured for ground and excited
states~\cite{M+P} and also from the more physical  charge and matter
distributions  evaluated for the ground state (1S) meson~\cite{GKPM}.
These results suggest that 2~fm is a sufficient size for quenched
evaluations, which is confirmed by   our results which extend the
spatial volume of the previous quenched measurements but do not show any
 statistically significant differences. 

 Dynamical fermion configurations are more  sensitive to  finite size
effects since more loop effects are included, in particular pion
exchange around the boundary becomes important. The leading
correction~\cite{luscher} for the ground state is a relative energy 
shift of order $c e^{-mL_S}$, where $m$ is the pseudoscalar mass and $c$
a coefficient given by the $B^* B \pi$ coupling. For the excited 
states, the possibility of the decay to (or mixing with)  nearby
two-body energy levels  becomes relevant. The only excited state that 
couples to a low-lying two-body energy level is for the P$_-$ which has 
a mixing with $B\pi$  where the pion has momentum zero. Thus we expect 
an enhanced finite size effect may arise for P$_-$. We investigate this 
by using two spatial sizes (called DF1 and DF2, with $L_S$ of 1.7~fm and
2.3~fm, corresponding  to $m_{\pi}L_S$ = 6.7 and 9.0 respectively). We
see some sign of a shift  for  D$_+$  and F but it is not very
significant statistically.  

 Our data set with the finest lattice spacing has a relatively small 
volume (1.6~fm with $m_{\pi}L_S$=4.5, 6.5) and, for this situation, 
some  evidence of finite size effects for the nucleon has been  
presented~\cite{JLQCD}.  Some of our results  from this finest lattice
spacing are significantly different from the larger volume results 
described above. Since the order $a$ formalism used is different (NP
clover rather than tadpole-improved clover) we cannot select between
finite size effects  or lattice artifacts (order $a$ \textit{etc.}) as a
cause.   We would need to use a larger spatial volume at  these
parameters to evaluate finite size effects fully.

 (ii) {\em Quark mass dependence}. Let us first discuss the dependence
on the light valence quark mass. Experimental data on the heavy light
mesons with  $c$ or $b$ quarks suggest that there is little quark mass
dependence of  excitation energies (\textit{i.e.} energy differences to
the ground state pseudoscalar meson) when going from strange quarks to
lighter quarks.  For instance the mass splittings  $D^*(1^-)-D$ and
$D^*_s(1^-)-D_s$ are 141 and 144 MeV  respectively, while $D^*(2^+)-D$
and $D^*_s(2^+)-D_s$  are 593 and 604 MeV respectively~\cite{PDG}.

 We can also explore this on a lattice and quark masses (characterised 
by $[r_0 m(0^{-+})]^2$ where $3.4$ corresponds to strange quarks as 
discussed below)  are varied in the range  of 0.6 to 1.5 times the
strange quark.  This is shown for the P-wave states in
Fig.~\ref{bsjpmass_fig} which confirms that  there is no significant
slope. This means that interpolation to the strange quark mass is not
delicate in any way.  Extrapolation  to light valence quarks is less
straightforward and  one issue that must  be addressed is that some of
the excited heavy-light mesons are unstable to strong decay. Since an
excited L$_{\pm}$ state will have  a decay to $B\pi$ with angular
momentum given by L$\pm 1$, only the P$_-$ state can decay in an S-wave
which then gives the lowest threshold energy on a lattice because the
pion  can have momentum zero.   In our case, because of the discrete
momentum and unphysical light-quark mass values, we do not have any open
decay channels in our lattice evaluation, but they will  open when
extrapolating in light quark mass and to large spatial volume. 

 The issue of the extrapolation in the sea quark mass is difficult  to
resolve. We cover the range from no sea quarks (\textit{i.e.} quenched)
to  $N_f=2$ flavours of sea quark with mass corresponding to 0.6 times
the  strange quark. The evaluation with even lighter sea quarks is
computationally  too demanding.

(iii) {\em The continuum limit}. It is feasible to study the continuum
limit in quenched studies,  but for dynamical fermions we only have
access to a relatively narrow range of  lattice spacing ($a$ from 0.15~fm
to 0.1~fm). To make best use of this  limitation, we use an order $a$
improved clover formulation of the fermion action. The coarser lattice 
has a tadpole-based improvement coefficient while the finer lattice uses
a non-perturbatively improved  value. Because of this difference in
formalism, it is  not straightforward to extrapolate from these two data
sets to the continuum limit. We take this into account in assigning
errors.

 \medskip
 
 {\em Lattice spectrum}.

 We average the values discussed above of the various excitation
energies,  weighting relatively more small lattice spacing, large volume
and quark masses close to strange. Thus we obtain $r_0 \Delta E$ of
1.07(7) for P$_-$; 1.33(13) for P$_+$.  The next excited level is the 2S
 which is at 1.25(-13,+50), this is an average based on the larger
volume studies but with the error reflecting our results at
finer lattice spacing.
  For the D-waves there is also a  large spread so we quote  a range:
for  D$_-$ from 2.2 to 3.1, while for D$_+$ from 2.2 to 3.5. For the 
F-wave, we only have an operator which excites  both F$_+$ and F$_-$ so
that our result is for an average of these two states, with an
excitation energy around 3.4 to 4.4.

 We need a value of the scale $r_0$ appropriate to  light quark
spectroscopy, since the dynamics of the light quark is the main aspect
of heavy-light mesons. Thus we do not  use values of $r_0$ from
heavy-heavy studies (which tend to give  somewhat smaller values) but an
average of those from light-light  mesons which span the range of 0.5 to
0.55 fm, namely 0.525$\pm$0.25,  for a discussion see Ref.~\cite{dmm}. 
This value of $r_0$, combined with the estimate of the mass of the 
pseudoscalar meson made from strange quarks~\cite{cmcm} of 687 to 
695 MeV yields $r_0 m_{\pi} \approx 1.84$ which sets the scale for the 
strange quark.

In our application to  the heavy-light mesons with $N_f=2$ flavours of
sea-quark, we have used  valence quarks identical to the sea-quarks,
which is the case  where the theory is fully unitary. This can be
interpreted in two ways --- firstly as applying to  the spectrum  of
excited $b \bar{n}$ states (where $n$ is $u$ or $d$) with quark masses
heavier than the physical values.  Indeed in the Tables we give the
pseudoscalar meson mass  obtained by combining these light quarks.
 On the other hand, for our application to  the $b \bar{s}$ system, we
have also used  valence quark masses identical to the sea-quark mass. In
the real world, however, there is only one  flavour of strange quark, so
we can interpret our results as from one  flavour of strange valence
quark propagating in a sea with two flavours of  light quarks whose mass
happens to correspond to the  valence quark mass.   This is effectively
treating the strange quark as partially quenched  and further studies
would be needed to  treat fully all three flavours  of light quark in
the sea.

 In principle one can calculate corrections to the heavy quark limit
from the lattice, as discussed later. Here, however, we adopt a more
modest strategy and make partial  use of experimental data. 
 Thus to interpolate to $b$ quarks we combine our results in the static
limit  with experimental data~\cite{expt,PDG} for   the $c\bar{s}$
system  as shown in Fig.~\ref{qmfig}. For the P$_-$  state the
experimental excitation energy for charm quarks is 349 MeV while we
obtain for static quarks 404(31) MeV. Thus the interpolation to  $b$
quarks involves only small shifts - leading to 386(31) MeV. 
   This is close to the threshold for  decay emitting a kaon (a mass gap
of 404 MeV) and  probably  below it. So we do  expect this $b\bar{s}$ 
scalar meson to be very narrow, as was found for  the $c \bar{s}$
counterpart~\cite{expt}. The associated axial meson at 434(31) MeV above
the $B_s$ will be close to the $B^* K$ threshold (at 450 MeV) and should
also be very narrow. The P$_+$ states lie above the $BK$ threshold  but
since these states decay in a D-wave, the centrifugal barrier  effects
may cause them to have narrow  widths.

 For the 2S, D, F   states, we do not have any $c\bar{s}$ counterpart
available from experiment to allow this interpolation. Assuming,
however, that the slopes versus $m_c/m_Q$  are similar to those for the
P-wave case, then the static energy  values will be a good approximation
to those for $b$ quarks. Again the 2S pseudoscalar (and vector) states
could be sufficiently light that they lie close to the  $B^* K$ ($BK$ for
vector) threshold at 450 MeV (404MeV ) and so are  very narrow.

 The only experimental observation~\cite{PDG} of an excited  $B_s$ state
is the $B_s$(5850) which lies 483 MeV heavier than the $B_s$  and has a
width of 47(22) MeV. This mass value is indeed  in the region where we
predict a rich spectrum of excited $b\bar{s}$ states.

 As we see no sign of a significant light quark mass dependence in our 
excitation energies, we can use our results to predict the spectrum  of 
excited $b \bar{n}$ states albeit with a somewhat  larger systematic
error from the extrapolation to light quarks which we are assuming to be
a constant.
 The only experimental observation~\cite{PDG} of an excited $B$ state is
the $B^*_J$(5732) which lies 419 MeV heavier than the $B$ and has a width
of 128(18) MeV. This may indeed be composed of several states. The mass
value is indeed  in the region where we predict a rich spectrum of
excited $b\bar{n}$ states, even though they should not be especially
narrow since the  $B\pi$ and $B^* \pi$ decay channels are open.

As well as predicting the spectrum, lattice methods can be used to 
evaluate decay amplitudes~\cite{rhodecay} and this is feasible for 
the heavy-light systems too~\cite{prog}.

\section{Discussion}

 We first discuss the issue of the theoretical relationship  between the
static limit and realistic heavy quarks. Then we can use this 
discussion to organise  our comparison with other lattice determinations
of the heavy-light spectrum.

A precise description of heavy-light mesons is provided by the heavy
quark effective theory. The leading ($1/m_Q$)
corrections~\cite{Manohar:dt} to the static (\textit{i.e.} heavy quark) limit
arise from two sources:  kinetic and magnetic terms. The magnetic
contribution  splits each static energy level H  (with total light quark
angular momentum $j_q = L \pm {1 \over 2}$, called L$_{\pm}$ above) into
two with  total angular momentum $j_1 = j_q +{1 \over 2}$ and $j_2 = j_q
-{1 \over 2}$. They have masses given by

 \begin{eqnarray} 
 m_{H_1} &  = & m_{Q} + \Lambda_{H}  +\frac{\lambda_{H,K}} { 2 m_Q } +  
 ( 2 j_{1} + 1)   \frac{ \lambda_{H,B}} { 2 m_{Q} } \\
 m_{H_2} &  = & m_{Q} + \Lambda_{H}  +\frac{\lambda_{H,K}} { 2 m_Q } - 
 ( 2j_{2} + 1) \frac{ \lambda_{H,B} } {2 m_{Q} },
  \label{eq:master} 
 \end{eqnarray} 

 \noindent  where $m_Q$ is the mass of the heavy quark and $\Lambda_{H}$
is the binding energy. On the lattice there is a self-energy term
proportional to $1/a$, but  as we only discuss mass differences, this
will cancel.

 Here $\lambda_{H,K}$ arises from the insertion of the heavy quark kinetic
energy for state H \textit{i.e.}
 \begin{equation} 
 \lambda_{H,K} =  \langle H \mid \overline{Q} D^2_{T} Q \mid H \rangle.
 \end{equation} 

 \noindent  As the transverse kinetic energy is expected to be positive,
this implies that $\lambda_{H,K}$ should be positive also. However, it
is the difference of kinetic energies  between states that we need. In a
simple approach with a confining  potential, the  excited state would
have larger kinetic energy than a ground state, so the mass differences
between  the P- and S-wave states would increase as $m_Q$ is decreased, but
this is only a qualitative indication.

The coefficient $\lambda_{H,B}$ arises for state H from the insertion of
the $\bm{\sigma.B}$ term, where $\bm{\sigma}$ is the heavy quark spin 
and $\bm{B}$ is the chromomagnetic  field from the light quark. 
For the S-wave states ($B^{\star}$, $B$) , the $\lambda_{S,B}$ 
parameter can be estimated from the experimental $B^{\star}$ to $B$ mass
splitting

 \begin{equation}
 \lambda_{S,B} \sim \frac{1}{4} (M_{B^\star}^2 - M_{B}^2 )
 = 0.12 \mbox{GeV}^2.
 \end{equation}

 The NRQCD lattice formalism allows these $1/m_Q$ expressions to be
evaluated. The results from several recent
studies~\cite{collins,Hein,lewis} show that  essentially all excitation
energies increase as $m_Q$ is decreased  from the static limit. This is
what would be expected from the  kinetic energy correction above. A
note of caution, however, is that  the magnetic contribution from these
studies can be compared with  experimental data on the $B^{\star}$, $B$
splitting and underestimates it  by almost a factor of two. This
suggests that the NRQCD, as currently implemented,  is not reproducing
the magnetic contribution accurately. Thus predictions from NRQCD of 
hyperfine splitting may be underestimated.  In the NRQCD method one does
not take a continuum limit, but the approach  can be systematically
improved by including  more terms in the effective action and by
computing the coefficients of these terms (such as $\bm{\sigma. B}$)
non-perturbatively, although this is yet to be carried out to a level
such that systematic errors on the hyperfine splittings can be established.

 Another way to estimate the $1/m_Q$ corrections from lattice studies is
to  compare static results, such as ours, with results from 
relativistic propagating quarks, where a continuum limit may be taken.
Here recent results for charm quarks~\cite{dmm} do give a spectrum of $c
\bar{s}$ mesons  substantially in agreement with experiment  and hence
support the pattern of $1/m_Q$ corrections we show in Fig.~\ref{qmfig}
Note that Bali~\cite{Bali} used  an estimate of $1/m_Q$ effects  by 
taking the difference of the  quenched results in the static limit 
(from Michael and Peisa~\cite{M+P}) and  for charm (from
Boyle~\cite{boyle}) and using this to correct the $N_f=2$ static result
from SESAM~\cite{Bali}. This different procedure explains why  his
result for the scalar P-wave meson is heavier  than ours (by $2 \sigma
$) even though  he obtains a similar value for the P$_{-}$ energy
excitation  in the static limit.

We illustrate some of the above discussion by presenting a compilation 
of relevant lattice results in Fig.~\ref{lattfig}.
 Some older lattice calculations of the mass spectrum of P-wave 
heavy-light mesons have been reviewed recently~\cite{Bali,dmm}.
  Improved lattice calculations with reduced systematic and  statistical
errors are required to get  definitive
answers~\cite{diPierro:2003iw,Kronfeld:2003sd}.

Having  discussed the heavy quark effective theory,  we now discuss
the implications of our results for other models of  heavy-light mesons.

 A traditional way to understand such spectra would be  using a quark
model with an underlying potential description~\cite{poker}. This  is
not strictly justified for a light quark, but may be of qualitative use.
For the experimentally observed excited $D_s$ states, it is difficult to
understand why the hyperfine splitting is sufficiently big to give a
$J^P=0^+$ meson which is so light in such an approach~\cite{bcl}.
 Our results enable us to discuss the possible  inversion of the level
ordering (with L$_+$ lighter than L$_-$)  at larger L or for radial
excitations.  This inversion  has been predicted~\cite{HJ1,HJ2}  from
consideration of the spin-orbit force, which at larger separation would
come   more from the confining interaction than the short-ranged
contribution from  gluon exchange. 
 We find no evidence of a sign change in the spin-orbit  splitting of
 P- or   D-waves. Thus conventional short-distance spin-orbit  effects are
still relevant up to radii appropriate for D-wave states.

 It is possible to discuss chiral symmetry in the heavy quark limit. 
 This allows relationships~\cite{chiral} between energy levels and also
predictions  for coupling strengths.  A stronger assumption of the form
of chiral  symmetry breaking allows to obtain results away from the
static limit, such as that  the $1^-$ to $0^-$ splitting is the same as
that from $1^+$ to $0^+$.   Chiral symmetry in the heavy quark limit  
relates the S  state to P$_-$  and the  D$_-$ state to P$_+$, \textit{etc.}
This does not seem to be a very good  approximation:  the spectrum is closer
to being dependent on L alone.  Indeed our spectrum shows an approximately
linear rise in excitation energy with L.

\begin{table}
 \begin{tabular}{cc}
 \hline                                                                    
   $J^P$         & $M(B_s^{*})-M(B_s)$ MeV \\
 \hline
  $0^+$          &  386$\pm$31 \\
  $1^+$          &  434$\pm$31 \\
  $1^+$          &  522$\pm$52 \\
  $2^+$          &  534$\pm$52 \\
  $0^-$          &  470{+188}{-52} \\
  $1^-$          &  470{+188}{-52}  \\
 \hline 
\end{tabular}	    

 \caption{ Lattice results for the energies of $b \bar{s}$ orbital (L=1)
and radial (2S) excited states. }
  \label{tab_sum}
\end{table}

\section{ Conclusion}

 We have used lattice QCD to explore the spectrum of heavy-light mesons.
 Our results are evaluated for static quarks but they are very relevant
to $b$ quarks ($B^*$ states) as we have argued.  We have concentrated
our studies on light quarks which are of the mass of a strange quark,
although we find that excitation energies are consistent with being
independent of the light quark mass, and hence will apply also to light
quarks which are $u$ and $d$. 

 In our lattice studies, we have pushed towards light sea quarks, towards 
small lattice spacing and towards large volume, but not 
towards all three requirements simultaneously. This leaves some room for 
systematic errors in our predictions. These can be reduced by further 
studies with increased computational resources. 

 We have determined the spectrum up to F-waves and including  radial
excitations, see Fig.~\ref{spdffig}.  This gives a rich texture for
model building of  heavy-light mesons. We find no evidence of a sign
change in the spin-orbit  splitting of  D-waves. Thus conventional
short-distance spin-orbit  effects are still relevant up to radii
appropriate for D-wave states. Rather than the pattern given by chiral
symmetry (which  relates S to P$_-$  and D$_-$ to P$_+$
\textit{etc.}~\cite{chiral}) we find a spectrum which is closer to being
dependent on L alone.  Indeed we see an approximate linear rise in
excitation energy with L, up to L=3, as 0.45 L GeV, reminiscent of Regge
or string models.

 We have discussed corrections to the heavy quark limit appropriate 
to $B^*$ states and have used experimental data on $D^*$ states 
to establish this.  
 Our results for the P-wave excitations confirm those obtained
previously~\cite{M+P} that the excitation energies are relatively small.
 This implies that the $b \bar{s}$ P-wave states will be close to the
lightest hadronic decay thresholds (namely $BK$ and $B^* K$).  The P$_-$
states ($J^P =0^+,\ 1^+$) have an S-wave decay but are light enough that
there is little or no phase space for decay.  The P$_+$ states ($J^P
=1^+,\ 2^+$) are heavier but have D-wave decays and so will also have
narrow widths since centrifugal barrier effects will reduce them.  We
also see evidence that the 2S states ($J^P =0^-,\ 1^-$) are close to the
lightest thresholds and so may be narrow too.  Our results for $b
\bar{s}$ states are summarised in Fig.~\ref{qmfig} and in 
Table~\ref{tab_sum}.  This raises the prospect that there will be many
(up to 6) narrow excited $B_s$ states to be found experimentally. 

Since we see no significant dependence of the excitation energies on
light quark mass, our predictions from the Table~\ref{tab_sum} can also
be used as estimates for orbital and radial $B^{*}-B$ excitation
energies, although these mesons will not be especially narrow since the
$B \pi$ and $B^* \pi$ thresholds are open. 

\vspace{1cm}
{\bf Acknowledgments}
\vspace{5mm}

The authors wish to thank the Center for Scientific Computing in Espoo, 
Finland and the ULGrid project of the University of Liverpool for making
available ample computer resources and thank  Chris Maynard for helpful
advice.  Two of us, AMG and JK, acknowledge  support by the Academy of
Finland contract 54038 and useful discussions with Petrus Pennanen. JK
thanks the Finnish Cultural Foundation and the Magnus Ehrnrooth
Foundation for financial support.

\begin{figure}[ht] 

\begin{center}
\includegraphics[angle=270,width=13cm]{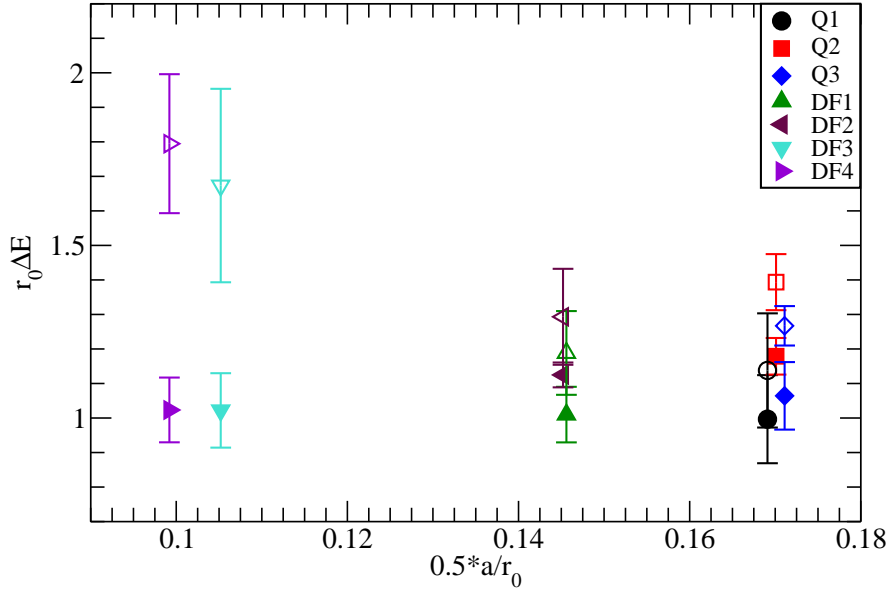}
\caption{The energies in units of $r_0$  of the  P$_+$ (open symbols)
and  P$_-$ (filled symbols)  levels with respect to the 1S energy for
different lattice spacings (approximately in fermis with our preferred 
value of $r_0=0.525$ fm).
  } 
\label{bsjp_fig} 
\end{center}

\end{figure}

\begin{figure}[ht] 

\begin{center}
\includegraphics[angle=270,width=13cm]{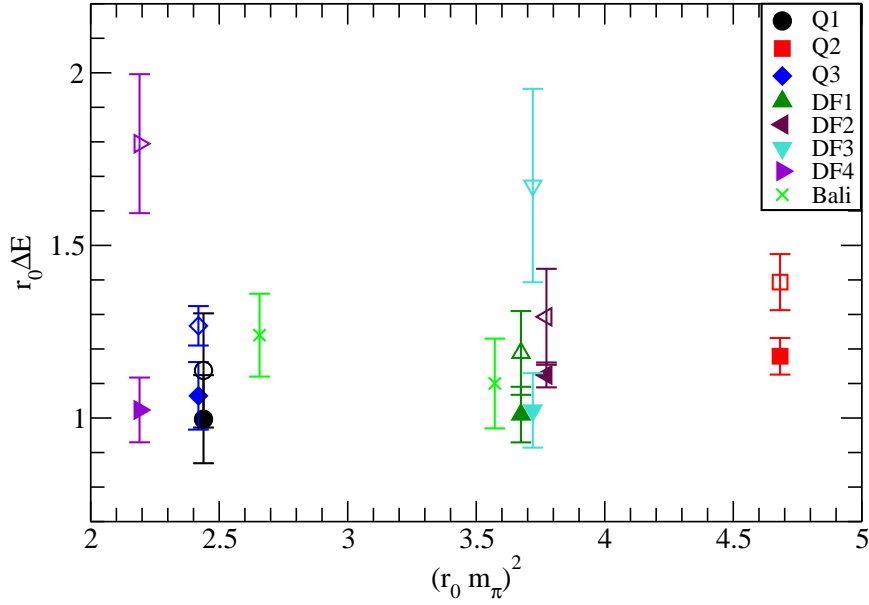}
\caption{The energies in units of $r_0$  of the  P$_+$ (open symbols)
and  P$_-$ (filled symbols)  levels with respect to the 1S energy for
different quark masses (as given by $[r_0 m(0^{-+})]^2$). Strange quarks 
correspond to a value of about 3.4.
  } 
\label{bsjpmass_fig} 
\end{center}

\end{figure}

\begin{figure}[ht] 

\begin{center}
\includegraphics[angle=270,width=13cm]{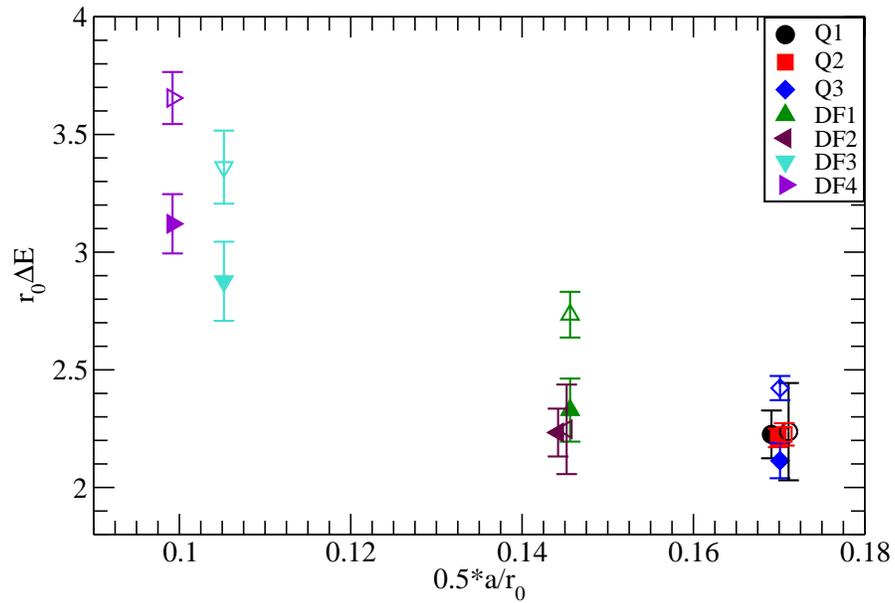}
\caption{The energies in units of $r_0$  of the  D$_+$ (open symbols)
and  D$_-$ (filled symbols)  levels with respect to the 1S energy for
different lattice spacings. 
  } 
\label{bsjd_fig} 
\end{center}

\end{figure}

\begin{figure}[ht] 

\begin{center}
\includegraphics[angle=270,width=13cm]{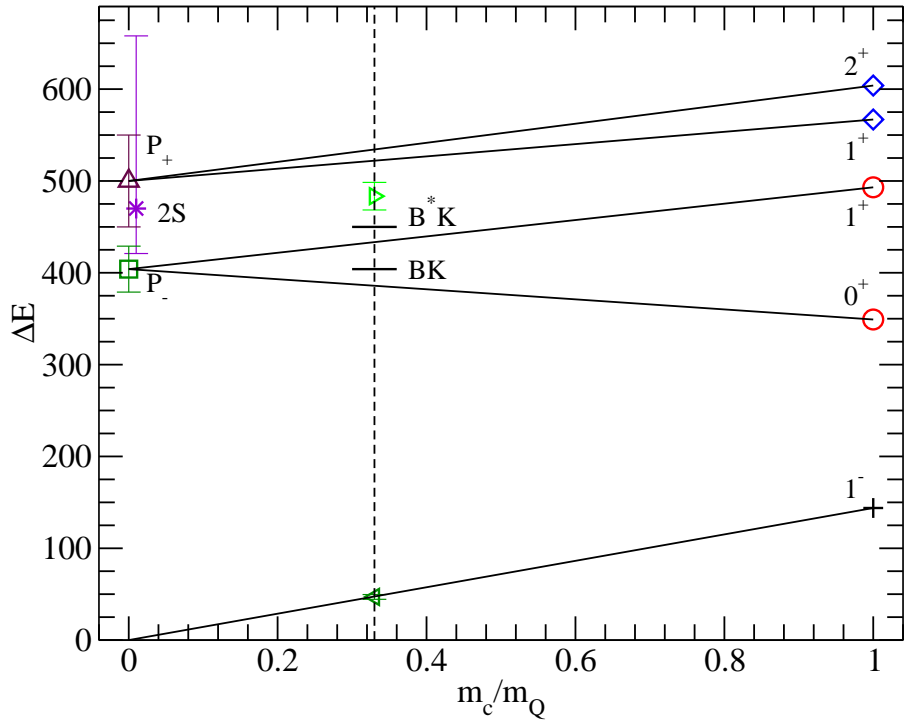}
 \caption{The energies in MeV of P-wave excited states relative to the
ground state ($J^P=0^-$) heavy-light  meson with heavy quark mass $m_Q$
and light quark which is strange. Data from experiment  are plotted for
charm and for $b$ quarks while our lattice results  are shown for static
quarks.  The 2S  excitation (from our larger volume results) is also
shown. The dotted vertical line gives the interpolated  value
appropriate for $b$ quarks.  The $BK$  and $B^* K$ thresholds are also
shown. These are the lightest isospin-conserving  decay modes allowed by
strong interactions.
  } 
\label{qmfig} 
\end{center}

\end{figure}

\begin{figure}[ht] 

\begin{center}
\includegraphics[angle=270,width=13cm]{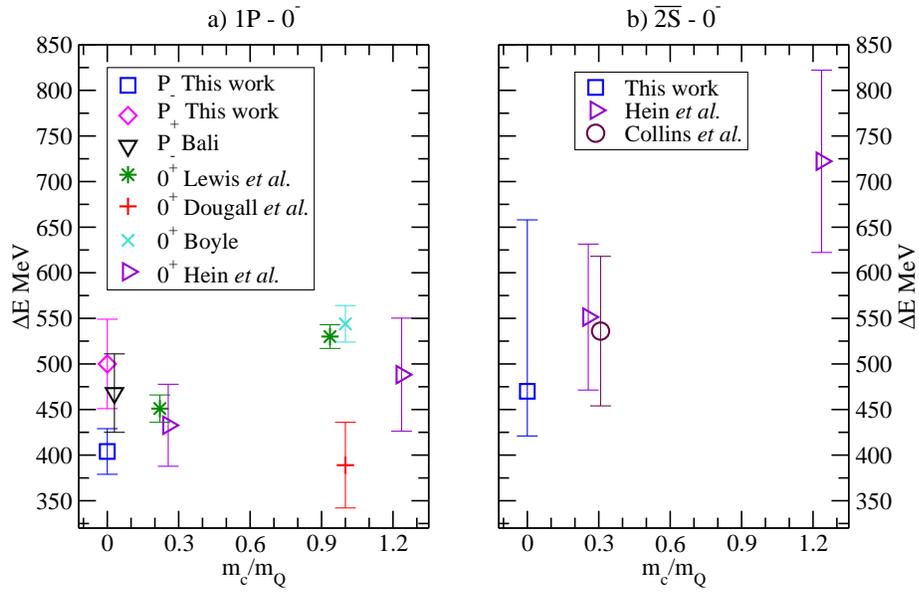}
 \caption{The energies from  lattice studies   of heavy-light excited
states  relative to the ground state ($J^P=0^-$) heavy-light  meson with
heavy quark of  mass $m_Q$ and light quark which is strange.  For clarity
we have displaced some of the numbers on the x axis, the graph should be
viewed as three clumps of numbers with heavy quarks at static, bottom
and charm.
  } 
\label{lattfig} 
\end{center}

\end{figure}

\begin{figure}[ht] 

\begin{center}
\includegraphics[angle=270,width=13cm]{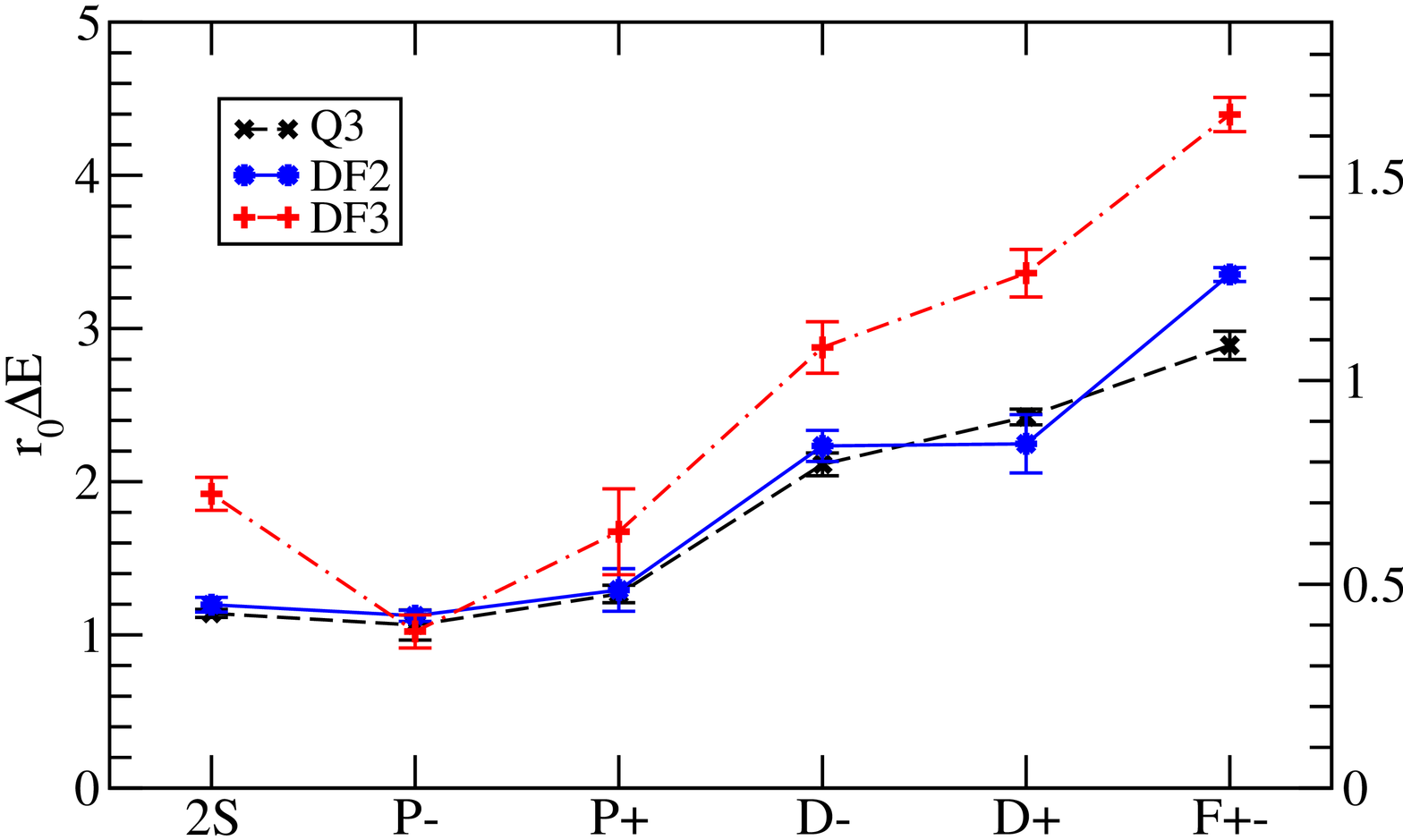}
 \caption{The energies from some of our lattice studies with $N_f=2$
(DF)  and $N_f=0$ (Q)  in units of $r_0$  of L-wave excited states and
the S-wave radial excited state relative to the ground state (1S)
heavy-light  meson with static heavy quark  and light quark which is
strange. 
  } 
\label{spdffig} 
\end{center}

\end{figure}


\begin{thebibliography}{99}

\bibitem{Rosner:2003cz}
J.~L.~Rosner,
hep-ph/0305315.

\bibitem{Gronau:1992ke}
M.~Gronau, A.~Nippe and J.~L.~Rosner,
Phys.\ Rev.\ D {\bf 47}, 1988  (1993).

\bibitem{M+P} C. Michael and J. Peisa  [UKQCD Collaboration], Phys. Rev. D
{\bf 58},  034506 (1998).

\bibitem{Bali} 
G.~S.~Bali,
Phys.\ Rev.\ D {\bf 68},  071501 (2003).

\bibitem{collins} S. Collins \textit{et al}.,
Phys.\ Rev.\ D\ {\bf 60}, 074504 (1999).

\bibitem{Hein} J. Hein \textit{et al}.,
Phys.\  Rev.\ D\ {\bf 62}, 074503  (2000).

\bibitem{lewis}
R.~Lewis and R.~M.~Woloshyn,
Phys.\ Rev.\ D {\bf 62}, 114507  (2000).

\bibitem{expt} BABAR Collaboration, B. Aubert \textit{et al}.,
Phys.\ Rev.\ Lett.\ {\bf 90}, 24200 (2003);
CLEO Collaboration, D. Besson \textit{et al}., hep-ex/0305017;
BELLE Collaboration, K. Abe \textit{et al}., hep-ex/0307021.

\bibitem{HJ1} H. J. Schnitzer, Phys.\ Rev.\ D {\bf 18}, 3482 (1978).

\bibitem{HJ2} H. J. Schnitzer, Phys. Letts.  {\bf B226}, 171 (1989).


\bibitem{chiral}   W. A. Bardeen, E. J. Eichten and C. T. Hill,
hep-ph/0305049; M. Novak, M. Rho and I. Zahed, Phys.\
Rev.\ D {\bf 68}, 054024 (2003).


\bibitem{Allton176} C.  R.  Allton \textit{et al}. [UKQCD Collaboration],
Phys.\
Rev.\ D {\bf 60}, 034507 (1999). 

\bibitem{Allton202} C.  R.  Allton \textit{et al}. [UKQCD Collaboration],
Phys.\ Rev.\ D {\bf 65}, 054502 (2002). 


\bibitem{GKPM}   A.M. Green, J. Koponen, P. Pennanen and C. Michael 
[UKQCD Collaboration], Eur.\ Phys.\ J.\ C {\bf 28}, 79 (2003).

\bibitem{luscher} M. L\"uscher, Comm.\ Math.\ Phys.\ {\bf 104}, 177 (1986).

 \bibitem{JLQCD} JLQCD Collaboration, S. Aoki \textit{et al}.,
Phys.\ Rev.\ D {\bf 68}, 054502 (2003).

\bibitem{PDG} K. Hagiwara \textit{et al}.,
Phys.\  Rev.\ D {\bf 66}, 010001 (2002).

\bibitem{dmm} A. Dougall, R. Kenway, C. Maynard and C. McNeile
[UKQCD Collaboration], Phys. Letts.  {\bf B569}, 41 (2003).


\bibitem{cmcm} C. McNeile and C. Michael  [UKQCD Collaboration], 
Phys.\ Lett.\ B {\bf 491}, 123  (2000).


\bibitem{rhodecay} C.  McNeile and C.  Michael [UKQCD Collaboration],
Phys.\ Lett.\ B {\bf 556}, 177 (2003). 


\bibitem{prog} C. Michael and G. Thompson (in preparation).

\bibitem{Manohar:dt}
A.~V.~Manohar and M.~B.~Wise,
Cambridge Monogr.\ Part.\ Phys.\ Nucl.\ Phys.\ Cosmol.\  {\bf 10}, 1 (2000).

\bibitem{boyle}
P.~Boyle  [UKQCD Collaboration],
Nucl.\ Phys.\ Proc.\ Suppl.\  {\bf 53}, 398 (1997).

\bibitem{diPierro:2003iw}
M.~di Pierro \textit{et al}., hep-lat/0310045.


\bibitem{Kronfeld:2003sd}
A.~S.~Kronfeld, hep-lat/0310063.


\bibitem{poker} A. M. Green, M. Jahma, J. Koponen and J. Ignatius (in
preparation).


\bibitem{bcl} T. Barnes, F. E. Close and H. J. Lipkin,
Phys.\  Rev.\ D {\bf 68}, 054006 (2003).


\end{thebibliography}
\end{document}